\documentclass[preprint,showpacs,preprintnumbers,amsmath,amssymb,natbib]{revtex4}

\usepackage{graphicx}
\usepackage{epsfig}		
\usepackage{dcolumn}
\usepackage{bm}
\usepackage{amsmath}
\usepackage{braket}
\usepackage{natbib}

\def\0{\mbox{\tiny $0$}}
\def\1{\mbox{\tiny $1$}}
\def\2{\mbox{\tiny $2$}}
\def\3{\mbox{\tiny $3$}}
\def\4{\mbox{\tiny $4$}}
\def\5{\mbox{\tiny $5$}}
\def\6{\mbox{\tiny $6$}}
\def\7{\mbox{\tiny $7$}}
\def\8{\mbox{\tiny $8$}}
\def\9{\mbox{\tiny $9$}}
\def\a{\mbox{\tiny $\alpha$}}

\def\f14{\mbox{\tiny $\frac{1}{4}$}}

\def\a{\mbox{\tiny $a$}}

\begin{document}

\title{Aspects of Phase-Space Noncommutative Quantum Mechanics}

\author{O. Bertolami}
\altaffiliation{Also at Centro de F\'isica do Porto, Rua do Campo Alegre 687, 4169-007, Porto, Portugal.}
\email{orfeu.bertolami@fc.up.pt}
\affiliation{Departamento de F\'isica e Astronomia, Faculdade de Ci\^{e}ncias da
Universidade do Porto, Rua do Campo Alegre 687, 4169-007, Porto, Portugal.}
\author{P. Leal}
\email{up201002687@fc.up.pt}
\affiliation{Departamento de F\'isica e Astronomia, Faculdade de Ci\^{e}ncias da
Universidade do Porto, Rua do Campo Alegre 687, 4169-007, Porto, Portugal.}


\begin{abstract}
In this work some issues in the context of Noncommutative Quantum Mechanics (NCQM) are addressed. The main focus is on finding whether symmetries present in Quantum Mechanics still hold in the phase-space noncommutative version. In particular, the issues related with gauge invariance of the electromagnetic field and the weak equivalence principle (WEP) in the context of the gravitational quantum well (GQW) are considered. The question of the Lorentz symmetry and the associated dispersion relation is also examined. Constraints are set on the relevant noncommutative parameters so that gauge invariance and Lorentz invariance holds. In opposition, the WEP is verified to hold in the noncommutative set up, and it is only possible to observe a violation through an anisotropy of the noncommutative parameters.
\end{abstract}

\date{\today}
\maketitle

\section{Introduction} \label{intro}

Noncommutative Quantum Mechanics is an extensively studied subject \cite{Gamboa:2001,Nair:2001,Duval:2001,Pei-Ming:2002,Horvathy:2002,Zhang:2004,Acatrinei:2005,Bastos:2006ps,2010CMaPh.299..709B,Bertolami:2005jw,Bernardini} and its interest arises for many reasons, more particularly from the fact that noncommutativity is present string theory and quantum gravity and black hole models (see e.g. \cite{SW,Connes,BastosQCBH}). NCQM can be viewed as the low-energy and the finite number of particles limit of noncommutative field theories and its main difference from standard quantum mechanics is the inclusion of an additional set of commutation relations for position and momentum operators. The Heisenberg-Weyl algebra for these operators,

\begin{equation} \label{heisenberg}
\left[\hat{x}_i,\hat{x}_j\right]=0, \hspace{40pt}
\left[\hat{p}_i,\hat{p}_j\right]=0, \hspace{40pt}
\left[\hat{x}_i,\hat{p}_j\right]=\mathrm{i}\hbar\delta_{ij}
\end{equation}
is deformed to the NC algebra:

\begin{equation} \label{NC algebra}
\left[\hat{q}_i,\hat{q}_j\right]=\mathrm{i}\theta_{ij}, \hspace{40pt}
\left[\hat{\pi}_i,\hat{\pi}_j\right]=\mathrm{i}\eta_{ij}, \hspace{40pt}
\left[\hat{q}_i,\hat{\pi}_j\right]=\mathrm{i}\hbar\delta_{ij},
\end{equation}
where $\theta_{ij}$ and $\eta_{ij}$ are anti-symmetric real matrices. The two sets of variables, $\{\hat{x}_i,\hat{p}_i\}$ and $\{\hat{q}_i,\hat{\pi}_i\}$ are related by a non-canonical linear transformation usually refered to as Darboux transformation, also known as Seiberg-Witten (SW) map. It is known that, although this map is not unique, all physical observables are independent of the chosen map \cite{Bastos:2006ps,2010CMaPh.299..709B}. Moreover, since the NC operators are defined in the same Hilbert space as the commutative ones, one can obtain a representation of them, up to some order of the noncommutative parameters, without the need for the Darboux transformation. However, in most cases, it is simpler to use this transformation in order to recover some known aspects of quantum mechanics. 

Besides the well-known operator formulation of quantum mechanics, a phase-space formulation of NCQM has been constructed \cite{Bastos:2006ps,2010CMaPh.299..709B} which allows for a straightforward implementation of noncommutativity. This formulation is useful for treating general problems such as, for instance, in cases where the potential is not specified. In this case, the position noncommutativity may be treated by a change in the product of functions to the Moyal $\star$-product, defined as:

\begin{equation}
A(x)\star_{\theta}B(x):=A(x)\mathrm{e}^{(\mathrm{i}/2)(\overleftarrow{\partial_{x_i}})\theta_{ij}(\overrightarrow{\partial_{x_j}})}B(x),
\end{equation}
and the momentum noncommutativity is introduced via a Darboux transformation. In the case of simple potentials, the use of the Darboux transformation ensures on its own, up to some order of the noncommutative parameter, a suitable noncommutative formulation.

Throughout the following sections, whenever need, the Darboux transformation to be used is as follows \cite{Bastos:2006ps}:

\begin{equation} \label{Darboux}
\hat{q}_i=\hat{x}_i-\frac{\theta_{ij}}{2\hbar}\hat{p}_j, \qquad \hat{\pi}_i=\hat{p}_i+\frac{\eta_{ij}}{2\hbar}\hat{x}_j.
\end{equation}

\section{Gauge Invariance}

In order to study the effects of NCQM we shall consider some physical systems of interest and investigate the implications of the NC deformation. The first example to consider is that of a particle with mass $m$ and charge $q$ in a magnetic field, with the Hamiltonian given by
\begin{equation}
\hat{H}= \frac{1}{2m}\left[\boldsymbol{\hat{\pi}}-q\boldsymbol{A}(\boldsymbol{q})\right]^2.
\end{equation}

In order to study this system we use the Moyal $\star$-product for the product of terms and then use the Darboux transformation, Eq. (\ref{Darboux}), to write the noncommuting Hamiltonian in terms of the commuting variables, $\hat{x}$ and $\hat{p}$. Thus, considering,
\begin{equation}
\hat{H}(\hat{q},\hat{\pi})\Psi(q)=\hat{H}(\hat{x},\hat{\pi})\star_\theta\Psi(x)=\hat{H}(\hat{x},\hat{\pi})\mathrm{e}^{(\mathrm{i}/2)(\overleftarrow{\partial_{x_i}})\theta_{ij}(\overrightarrow{\partial_{x_j}})}\Psi(x),
\end{equation}
at first order in the parameter $\theta$,
\begin{gather}
\left[\hat{H}(\hat{x},\hat{\pi})+\frac{\mathrm{i}\theta_{ab}}{2}\partial_a\hat{H}(\hat{x},\hat{\pi})\partial_b\right]\Psi(x)= \nonumber\\
=\left[\frac{1}{2m}\left(\boldsymbol{\hat{\pi}}^2-2q\boldsymbol{\hat{\pi}}\cdot\boldsymbol{A}(\boldsymbol{q})+q^2A^2(q)\right)+\frac{\mathrm{i}\theta_{ab}}{2}\partial_a\left(q^2A^2(x)-2q\boldsymbol{A}(x)\cdot\hat{\pi}\right)\partial_b\right]\Psi(x)
\end{gather}


If we now consider that $\theta_{ab}=\theta\epsilon_{ab}$, where $\epsilon_{ab}$ is the 2-dimentional antisymmetric symbol, the effective noncommutative Hamiltonian, at first order in $\theta$, becomes:
\begin{equation}
\hat{H}=\frac{1}{2m}\left(\boldsymbol{\hat{\pi}}^2-2q\boldsymbol{\hat{\pi}}\cdot\boldsymbol{A}(\boldsymbol{q})+q^2A^2(q)\right)+\frac{\mathrm{i}}{4m}\left[\nabla\left(q^2A^2(x)-2q\boldsymbol{A}(x)\cdot\hat{\pi}\right)\times\nabla\right]\cdot\boldsymbol{\theta}
\end{equation}
where $\boldsymbol{\theta}=\theta(1,-1,1)$. We now make use of the Darboux transformation, Eq. (\ref{Darboux}), in the momentum operator (which is now the only noncommutative operator in the Hamiltonian) to obtain:
\begin{multline} \label{eq ncemg}
\hat{H}=\frac{1}{2m}\left[\left(\hat{\boldsymbol{p}}-q\boldsymbol{A}(\boldsymbol{x})\right)^2-\frac{1}{\hbar}(\hat{\boldsymbol{x}}\times\hat{\boldsymbol{p}})\cdot\boldsymbol{\eta}-\frac{q}{\hbar}(\hat{\boldsymbol{x}}\times\boldsymbol{A}(\boldsymbol{x}))\cdot\boldsymbol{\eta}+\frac{1}{4\hbar^2}\eta^2\epsilon_{ij}\epsilon_{ik}\hat{x}_j\hat{x}_k\right] \\
-\frac{1}{4m\hbar}\left[\nabla\left(q^2A^2(\boldsymbol{x})-2q\boldsymbol{A}(\boldsymbol{x})\cdot\hat{\boldsymbol{p}}-\frac{q}{\hbar}(\hat{\boldsymbol{x}}\times\boldsymbol{A}(\boldsymbol{x}))\cdot\boldsymbol{\eta}\right)\times\hat{\boldsymbol{p}}\right]\cdot\boldsymbol{\theta},
\end{multline}
where, as in the case of $\theta$, $\boldsymbol{\eta}=\eta(1,-1,1)$. We aim now to see how a gauge transformation modifies the Hamiltonian and study the condition under which the Hamiltonian is gauge invariant. Gauge invariance must be imposed, otherwise a gauge change would lead to a modification of the system energy for the same physical configuration. For this purpose, we consider a gauge transformation to the vector potential $\boldsymbol{A}\rightarrow\boldsymbol{A}'=\boldsymbol{A}+\boldsymbol{\nabla}\alpha$, where $\alpha$ is a scalar function of position. Consider now the first set of terms in the Hamiltonian, Eq. (\ref{eq ncemg}). Under the stated transformation, we get:
\begin{equation}
\begin{split}
\frac{1}{2m}\left[\left(\hat{\boldsymbol{p}}-q\boldsymbol{A}(\boldsymbol{x})-q\boldsymbol{\nabla}\alpha\right)^2-\frac{1}{\hbar}(\hat{\boldsymbol{x}}\times\hat{\boldsymbol{p}})\cdot\boldsymbol{\eta}- \right. \\
\left. -\frac{q}{\hbar}(\hat{\boldsymbol{x}}\times\boldsymbol{A}(\boldsymbol{x}))\cdot\boldsymbol{\eta}-\frac{q}{\hbar}(\hat{\boldsymbol{x}}\times\boldsymbol{\nabla}\alpha)\cdot\boldsymbol{\eta}+\frac{1}{4\hbar^2}\eta^2\epsilon_{ij}\epsilon_{ik}\hat{x}_j\hat{x}_k\right].
\end{split}
\end{equation}

If we now change the wave function on which the Hamiltonian acts, to $\Psi=\mathrm{e}^{\mathrm{i}q\alpha/\hbar}\Psi'$, the first set of extra terms in Eq. (\ref{eq ncemg}) coming from the gauge transformation will be cancelled and so we may conclude that this set of therms is not problematic. However, this is not true for the second set of terms which is transformed to,

\begin{equation}
\left[\nabla\left(q^2(A(\boldsymbol{x})+\boldsymbol{\nabla}\alpha)^2-2q\boldsymbol{A}(\boldsymbol{x})\cdot\hat{\boldsymbol{p}}-2q\boldsymbol{\nabla}\alpha\cdot\hat{\boldsymbol{p}}-\frac{q}{\hbar}(\hat{\boldsymbol{x}}\times\boldsymbol{A}(\boldsymbol{x}))\cdot\boldsymbol{\eta}-\frac{q}{\hbar}(\hat{\boldsymbol{x}}\times\boldsymbol{\nabla}\alpha)\cdot\boldsymbol{\eta}\right)\times\hat{\boldsymbol{p}}\right]\cdot\boldsymbol{\theta}.
\end{equation}

If we now consider the wave function transformation, $\Psi=\mathrm{e}^{\mathrm{i}q\alpha/\hbar}\Psi'$, we verify that the gauge transformation is not cancelled due to the momentum operator outside the divergence acting on the exponential. Thus, the phase transformation that absorbs the gauge transformation terms in the first part of the Hamiltonian, Eq. (\ref{eq ncemg}), does not do so for the second set of terms. This comes from the fact that, in the first term, the change in $\boldsymbol{A}$ can be seen as a change in $\hat{\boldsymbol{p}}$, and a constant change in momenta can always be absorbed by a phase change. The same does not occur for the change in the second term, making it impossible to accommodate it into a change in phase. Therefore, in order to make the Hamiltonian gauge invariant, this term must vanish. To accomplish this for any $\boldsymbol{A}$, $\theta$ must vanish. This result is consistent to an explicit computation in the context of the Hamiltonian of fermionic fields \cite{Bertolami:2011rv}.

\section{Gravitational Quantum Well and the Equivalence Principle in NCQM}

A very interesting system to directly connect gravity to quantum mechanics is the gravitational quantum well \cite{Landau,LF,Nesvizhesky}. As we shall see, this connection can be used to constrain quantum measurements of gravity phenomena and to test the equivalence principle (see also Refs. \cite{Bertolami:2003,Bastos:2010au}). It is easy to show that this principle holds for usual quantum mechanics, in the sense that a gravitational field is equivalent to an accelerated  reference frame. We shall see that this also holds in the context of NCQM for isotropic noncommutativity parameters. In the following we shall study the noncommutative GQW \cite{Bertolami:2005jw} and its connection to accelerated frames of reference.

\subsection{Fock space formulation of NC Gravitational Quantum Well}

Let us consider the GQW in the context of NCQM.
To start with we review some aspects of the usual GQW in standard quantum mechanics. The Hamiltonian is given by:

\begin{equation} \label{QGWH}
\hat{H}=\frac{1}{2m}\hat{\boldsymbol{p}}^2+mg\hat{x}_i.
\end{equation}
for a particle with mass, $m$, in a gravitational field with acceleration, $g$, in the $x_i$ direction.

With the Fock space treatment in mind we define creation and annihilation operators for this Hamiltonian:
\begin{gather}
\hat{b}=\left(\frac{m^2}{\hbar^2g}\right)^{\frac{1}{3}}\left[\hat{x}+\frac{i}{2}\left(\frac{g^2\hbar}{m^4}\right)^{\frac{1}{3}}\hat{p}_x\right], \label{annihilation}\\
\hat{b}^\dagger=\left(\frac{m^2}{\hbar^2g}\right)^{\frac{1}{3}}\left[\hat{x}-\frac{i}{2}\left(\frac{g^2\hbar}{m^4}\right)^{\frac{1}{3}}\hat{p}_x\right], \label{creation}
\end{gather}
where the definition concerns only for the $x$ direction, as the $y$ component of the Hamiltonian is just that of a free particle. The normalization factors is chosen so that the operators $\hat{b}$ and $\hat{b}^\dagger$ are dimensionless. The Hamiltonian can then be rewritten as 
\begin{equation} \label{comm hamiltonian}
\hat{H}=K_1\left(\hat{\Gamma}_x+\hat{\Gamma}_y\right)+K_2\left(\hat{b}^\dagger_x+\hat{b}_x\right),
\end{equation}
where
\begin{gather}
\hat{\Gamma}_i=\hat{b}^\dagger_i\hat{b}_i+\hat{b}_i\hat{b}^\dagger_i-\hat{b}^\dagger_i\hat{b}^\dagger_i-\hat{b}_i\hat{b}_i, \\
K_1=\frac{1}{16}\left(\frac{\hbar^3m^2}{g}\right)^{2/3}, \\
K_2=\frac{mg}{2}\left(\frac{\hbar^2g}{m^2}\right)^{1/3}.
\end{gather}

Given the form of the Hamiltonian, it is evident that it is not diagonal in this representation, so it is not particularly useful for calculations of eigenstates and eigenvalues. This is expected from the usual solution to this problem, in which the energies involve the zeros of the Airy function, $Ai(x)$. We now examine the noncommutative Hamiltonian \cite{Bertolami:2005jw},

\begin{equation} \label{NCGQWH}
\hat{H}^{NC}=\frac{1}{2m}\left[\hat{p}_x^2+\hat{p}_y^2\right]+mg\hat{x}+\frac{\eta}{2m\hbar}(\hat{x}\hat{p}_y-\hat{y}\hat{p}_x)+\frac{\eta^2}{8m\hbar^2}\left(\hat{x}^2+\hat{y}^2\right);
\end{equation}
which is the equation of a particle under the influence of a gravitational field plus a fictitious ``magnetic field", $\overrightarrow{B_{NC}}=-(\eta/q\hbar)\overrightarrow{\mathrm{e}_z}$, plus an harmonic restoring force. Through the definitions, Eqs. (\ref{annihilation}) and (\ref{creation}), it can be rewriten it, up to first order in $\theta$ and $\eta$, as:
\begin{equation} \label{noncomm hamiltonian}
\hat{H}^{NC}=K_1\left(\hat{\Gamma}_x+\hat{\Gamma}_y\right)+K_2\left(\hat{b}^\dagger_x+\hat{b}_x\right)+\frac{i\eta}{4m\hbar^{\frac{2}{3}}}\left(\hat{b}^\dagger_y\hat{b}_x-\hat{b}^\dagger_x\hat{b}_y\right).
\end{equation}

It should be pointed out that this treatment considers only first order terms in either $\eta$ or $\theta$, although the latter does not show up in the Hamiltonian as its effect can be absorbed by a phase factor of the wave function. Noting the similarities between both commutative and noncommutative Hamiltonians, we might ask wether there is a transformation that can turn one into the other. That might be an interesting finding as, then, noncommutativity, at least for this system, could be regarded as a modification to the commutative case, and noncommutative eigenfunctions could be constructed using commutative ones, which are well known. Furthermore, it would make noncommutativity the result of a transformation of variables, and not a fundamental property of the system under study.  In order to pursue this analysis, we must introduce an operator transformation in which the new operators, $\hat{a}_i$ and $\hat{a}^\dagger_i$ for $i=x,y$, obey the same commutation relations as the original operators. Thus we define,

\begin{gather} \label{a operators}
\hat{b}_i:=\sum_{j=1}^2u_{ij}\hat{a}_j+s_{ij}\hat{a}^\dagger_j, \\
\hat{b}^\dagger_i:=\sum_{j=1}^2u^*_{ij}\hat{a}^\dagger_j+s^*_{ij}\hat{a}_j,
\end{gather}
where we impose the commutation relations

\begin{equation}
\left[\hat{a}_i,\hat{a}^\dagger_j\right]=\delta_{ij},
\end{equation}
and all the other commutation relations vanish. These conditions introduce a set of constraints on the parameters $u_{ij}$ and $s_{ij}$, namely:

\begin{gather}
\lvert u_{11}\lvert^2-\lvert s_{11} \lvert^2+\lvert u_{12}\lvert^2-\lvert s_{12}\lvert^2=1, \nonumber\\
\lvert u_{21}\lvert^2-\lvert s_{21} \lvert^2+\lvert u_{22}\lvert^2-\lvert s_{22}\lvert^2=1.
\end{gather}

Considering Eq. (\ref{comm hamiltonian}) in terms of operators $\hat{b}_i$ and $\hat{b}^\dagger_i$ and using the definitions, Eq. (\ref{a operators}), we get the Hamiltonian in terms of the operators $\hat{a}_i$ and $\hat{a}^\dagger_i$ as 

\begin{multline} \label{new hamiltonian}
\hat{H}=K_1\left[\gamma_1\hat{a}^\dagger_x\hat{a}_x+\gamma_1\hat{a}_x\hat{a}^\dagger_x+\gamma_2\hat{a}^\dagger_x\hat{a}^\dagger_x+\gamma^*_2\hat{a}_x\hat{a}_x+\gamma_3\hat{a}^\dagger_y\hat{a}_y+\gamma_3\hat{a}_y\hat{a}^\dagger_y+\gamma_4\hat{a}^\dagger_y\hat{a}^\dagger_y+ \right. \\
\left. +\gamma^*_4\hat{a}_y\hat{a}_y+2\gamma_5\hat{a}^\dagger_x\hat{a}^\dagger_y+2\gamma^*_5\hat{a}_x\hat{a}_y+2\gamma_6\hat{a}^\dagger_x\hat{a}_y+2\gamma^*_6\hat{a}^\dagger_y\hat{a}_x\right]+ \\
+K_2\left[\hat{a}^\dagger_x\left(u^*_{11}+s_{11}\right)+\hat{a}_x\left(s^*_{11}+u_{11}\right)+\hat{a}^\dagger_y\left(u^*_{12}+s{12}\right)+\hat{a}_y\left(s^*_{12}+u_{12}\right)\right],
\end{multline}
where, for simplicity, we have defined,

\begin{subequations}
\begin{equation}
\gamma_1:=\lvert u_{11}\lvert^2+\lvert s_{11}\lvert^2-u^*_{11}s^*_{11}-u_{11}s_{11}+\lvert u_{21}\lvert^2+\lvert s_{21}\lvert^2-u^*_{21}s^*_{21}-u_{21}s_{21},
\end{equation}
\begin{equation}
\gamma_2:=2u^*_{11}s_{11}-\left(u^*_{11}\right)^2-s_{11}^2+2u^*_{21}s_{21}-\left(u^*_{21}\right)^2-s_{21}^2,
\end{equation}
\begin{equation}
\gamma_3:=\lvert u_{12}\lvert^2+\lvert s_{12}\lvert^2-u^*_{12}s^*_{12}-u_{12}s_{12}+\lvert u_{22}\lvert^2+\lvert s_{22}\lvert^2-u^*_{22}s^*_{22}-u_{22}s_{22},
\end{equation}
\begin{equation}
\gamma_4:=2u^*_{12}s_{12}-\left(u^*_{12}\right)^2-s_{12}^2+2u^*_{22}s_{22}-\left(u^*_{22}\right)^2-s_{22}^2,
\end{equation}
\begin{equation}
\gamma_5:=u^*_{11}s_{12}+s_{11}u^*_{12}-u^*_{11}u^*_{12}-s^*_{11}s^*_{12}+u^*_{21}s_{22}+s_{21}u^*_{22}-u^*_{21}u^*_{22}-s^*_{21}s^*_{22},
\end{equation}
\begin{equation}
\gamma_6:=u^*_{11}u_{12}+s_{11}s^*_{12}-u^*_{11}s^*_{12}-s^*_{11}u^*_{12}+u^*_{21}u_{22}+s_{21}s^*_{22}-u^*_{21}s^*_{22}-s^*_{21}u^*_{22}.
\end{equation}
\end{subequations}

Comparing the Hamiltonian in Eq. (\ref{new hamiltonian}) to the one in Eq. (\ref{noncomm hamiltonian}), we can immediately set the conditions for the $\gamma_i$'s

\begin{subequations}
\begin{equation}
\gamma_1=1,
\end{equation}
\begin{equation}
\gamma_2=-1,
\end{equation}
\begin{equation}
\gamma_3=1,
\end{equation}
\begin{equation}
\gamma_4=-1,
\end{equation}
\begin{equation}
\gamma_5=0,
\end{equation}
\begin{equation}
\gamma_6=\mathrm{i}\frac{\eta}{4m\hbar^{\frac{2}{3}} K_1}:=i\eta\mathrm{c} , \mathrm{c}\in \mathbb{R}.
\end{equation}
\end{subequations}

Furthermore, comparing the terms that are linear in the $\hat{a}$ operators, we get two additional equations for the $u$ and $s$ parameters,

\begin{subequations}
\begin{equation}
u^*_{11}+s_{11}=1,
\end{equation}
\begin{equation}
u^*_{12}+s_{12}=0.
\end{equation}
\end{subequations}

In total we now have 16 variables and a total of 16 distinct equations constraining the values of this variables. Hence, this system of equations has either a single solution or none. It is found that this system has no solution for $\eta\neq 0$, which can be verified using well known Mathematica or MatLab procedures. Therefore, it is not possible to describe, as expected, the noncommutative Hamiltonian as a mixture of eigenstates of the commutative Hamiltonian, and so it is a completely different problem. Once again we stress that this result is only valid at first order in both noncommutative parameters. However, it is reassuring to confirm that, at least at this level, noncommutativity is indeed a completely different problem than the commutative one.

\subsection{Equivalence Principle}

Having verified that the noncommutative Hamiltonian of the GQW is in fact a different problem than the commutative one, we can try to examine the issue of the noncommutative Equivalence Principle. We have seen that the only parameter having an effect on the eigenstates and eigenvalues is $\eta$, as the $\theta$ factor can be absorbed into a phase factor in the wave function of the system. The WEP states that, locally, any gravitational field is equivalent to an accelerated reference frame. This is one of the basic tenets of General Relativity and holds with great accuracy (see e.g. Ref. \cite{Bertolami:2012}, chapter 22, for a review of the experimental status of relativity). In standard QM, for the GQW, this can be verified to hold in a quite simple way. In the context of NCQM we will show how it can be verified in what follows next. For this purpose we consider the noncommutative GQW Schr$\ddot{\mathrm{o}}$dinger equation,

\begin{equation}
\hat{H}^{NC}_g\Psi=\left[\frac{1}{2m}\left(\hat{\pi}_x^2+\hat{\pi}_y^2\right)+mg\hat{Q}_x\right]\Psi=E\Psi
\end{equation}
and applying the Darboux transformation to write it in terms of the commutative variables, that is, Eq. (\ref{NCGQWH}): 

\begin{equation} \label{schrodinger of NC GQW}
\left[\frac{1}{2m}\left(\hat{p}_x^2+\hat{p}_y^2\right)+mg\hat{x}+\frac{\eta}{2m\hbar}(\hat{x}\hat{p}_y-\hat{y}\hat{p}_x)+\frac{\eta^2}{8m\hbar^2}\left(\hat{x}^2+\hat{y}^2\right)\right]\Psi=\mathrm{i}\hbar\frac{\partial\Psi}{\partial t},
\end{equation}
where we have considered the time dependent problem as we have to use a change of coordinates evolving in time. We now consider the noncommutative free particle equation:

\begin{equation} \label{free hamiltonian}
\left[-\frac{\hbar^2}{2m}\left(\frac{\partial^2}{\partial x^2}+\frac{\partial^2}{\partial y^2}\right)-\frac{\mathrm{i}\eta}{2m}(x\frac{\partial}{\partial y}-y\frac{\partial}{\partial x})+\frac{\eta^2}{8m\hbar^2}\left(x^2+y^2\right)\right]\Psi=\mathrm{i}\hbar\frac{\partial\Psi}{\partial t},
\end{equation}
and introduce a change of coordinates defined as

\begin{subequations} \label{acc coordinates}
\begin{equation}
x'=x+\sigma(t)
\end{equation}
\begin{equation}
y'=y
\end{equation}
\end{subequations}

In order for the WEP to be preserved we require that

\begin{equation} \label{equality}
\hat{H}^{NC}_g(\hat{\boldsymbol{x}},\hat{\boldsymbol{p}})\Psi(x,y)=\hat{H}^{NC}_{free}(\hat{\boldsymbol{x'}},\hat{\boldsymbol{p'}})\Psi'(x',y'),
\end{equation}
where $\hat{H}^{NC}_g$ is the noncommutative GQW Hamiltonian and $\hat{H}^{NC}_{free}$ is the noncommutative Hamiltonian of a free particle and $\Psi'(x',y')=\mathrm{e}^{\mathrm{i}\phi(x',y')}\Psi(x',y')$, so that the eigenfunctions are the same, but by a phase. Starting from the free particle Hamiltonian we write it in terms of an accelerated reference frame coordinates, and thus,
\begin{subequations} \label{acc differentials}
\begin{equation}
\frac{\partial}{\partial x'}\Psi(x',y')=\frac{\partial}{\partial x}\Psi(x,y),
\end{equation}
\begin{equation}
\frac{\partial}{\partial y'}\Psi(x',y')=\frac{\partial}{\partial y}\Psi(x,y),
\end{equation}
\begin{equation}
\frac{\partial}{\partial t'}\Psi(x',y')=\left(\frac{\partial}{\partial t}-\frac{\mathrm{d}\sigma(t)}{\mathrm{d}t}\frac{\partial}{\partial x}\right)\Psi(x,y).
\end{equation}
\end{subequations}

Hence, combining Eqs. (\ref{acc coordinates}) and (\ref{acc differentials}), the right-hand side of Eq. (\ref{free hamiltonian}) becomes:
\begin{multline}
\left[-\frac{\hbar^2}{2m}\left(\frac{\partial^2}{\partial x^2}+\frac{\partial^2}{\partial y^2}\right)-\frac{\mathrm{i}\eta}{2m}\left(x\frac{\partial}{\partial y}-y\frac{\partial}{\partial x}\right)-\frac{\mathrm{i}\eta}{2m}\sigma(t)\frac{\partial}{\partial y}+\frac{\eta^2}{8m\hbar^2}\left(x^2+y^2\right)+ \right. \\
\left. \frac{\eta^2}{8m\hbar^2}\left(-2x\sigma(t)+\sigma^2(t)\right)\right]\Psi'(x,y)=\mathrm{i}\hbar\left(\frac{\partial}{\partial t}-\frac{\mathrm{d}\sigma(t)}{\mathrm{d}t}\frac{\partial}{\partial x}\right)\Psi'(x,y).
\end{multline}

In order to check if Eq. (\ref{equality}) is consistent we must either compute the phase $\phi$ or prove there is no wave function which holds for the mentioned relation. For this we consider the relation between $\Psi$ and $\Psi'$ and compute the action of the operators on the wave function $\Psi'(x',y')=\mathrm{e}^{\mathrm{i}\phi(x',y')}\Psi(x',y')$. The obtained result is as follows:
\begin{multline} \label{full equation}
\left[-\frac{\hbar^2}{2m}\left(\frac{\partial^2}{\partial x^2}+\frac{\partial^2}{\partial y^2}\right)-\frac{\mathrm{i}\eta}{2m}\left(x\frac{\partial}{\partial y}-y\frac{\partial}{\partial x}\right)+\frac{\eta^2}{8m\hbar^2}\left(x^2+y^2\right)\right]\Psi'+\left[-\frac{\mathrm{i}\hbar^2}{2m}\frac{\partial^2\phi}{\partial x^2}+\frac{\hbar^2}{2m}\frac{\partial\phi}{\partial x}^2 \right. \\
\left. -\frac{\mathrm{i}\hbar^2}{2m}\frac{\partial^2\phi}{\partial y^2}+\frac{\hbar^2}{2m}\frac{\partial\phi}{\partial y}^2+\frac{\eta}{2m}y\frac{\partial\phi}{\partial x}-\frac{\eta}{2m}x\frac{\partial\phi}{\partial y}+\frac{\eta}{2m}\sigma(t)\frac{\partial\phi}{\partial y}-\frac{\eta^2}{4m\hbar^2}x\sigma(t)+\frac{\eta^2}{4\hbar^2}\sigma^2(t)+ \right. \\
\left. \hbar\frac{\partial\phi}{\partial t}+\hbar\frac{\mathrm{d}\sigma}{\mathrm{d}t}\frac{\partial\sigma}{\partial x}\right]\Psi'+\left[-\frac{\mathrm{i}\hbar^2}{2m}\frac{\partial\phi}{\partial x}+\mathrm{i}\hbar\frac{\mathrm{d}\sigma}{\mathrm{d}t}\right]\frac{\partial\Psi'}{\partial x}+\left[-\frac{\mathrm{i}\hbar^2}{m}\frac{\partial\phi}{\partial y}-\frac{\mathrm{i}\eta}{2m}\sigma(t)\right]\frac{\partial\Psi'}{\partial t}=\mathrm{i}\hbar\frac{\partial\Psi'}{\partial t}.
\end{multline}

Now, for the purpose of retrieving the noncommutative GQW we must compare both Schr$\ddot{\mathrm{o}}$dinger equations to set constraints on the form of the phase $\phi$. Imposing that the term multiplying the derivative of $\Psi'$ vanishes, we get:
\begin{equation}
\frac{\partial\phi}{\partial x}=\frac{m}{\hbar}\frac{\mathrm{d}\sigma}{\mathrm{d}t},
\end{equation}
which implies, taking into account the fact that $\sigma$ only depends on time, that:
\begin{equation} \label{first form}
\phi=\frac{m}{\hbar}\frac{\mathrm{d}\sigma}{\mathrm{d}t}x+f(y,t).
\end{equation}

Considering that the last term on the left-hand side of Eq. (\ref{full equation}) must vanish, and Eq. (\ref{first form}), it follows that
\begin{equation}
\frac{\hbar^2}{m}\frac{\partial f}{\partial y}=-\frac{\eta}{2m}\sigma(t)\space\Rightarrow\space f(y,t)=-\frac{\eta}{2\hbar^2}\sigma(t)y+\mu(t);
\end{equation}
replacing this result into the second term of Eq. (\ref{full equation}) and comparing with the Hamiltonian, Eq. (\ref{schrodinger of NC GQW}), yields
\begin{equation}
m\frac{\mathrm{d}^2\sigma}{\mathrm{d}t^2}x+\nu(t)=mgx
\end{equation}
where $\nu(t)$ is the sum of all time dependent terms and can be made to vanish through a suitable choice of the function $\mu(t)$. There is only one non-vanishing remaining term and in order to Eq. (\ref{equality}) to hold we must impose that
\begin{equation}
\frac{\mathrm{d}^2\sigma}{\mathrm{d}t^2}=g\space\Rightarrow\space\sigma(t)=\sigma_0+vt+\frac{1}{2}gt^2
\end{equation}

Thus, we can see that Eq. (\ref{equality}) holds as far as
\begin{equation}
x'=x+\sigma_0+vt+\frac{1}{2}gt^2
\end{equation}
which corresponds to an accelerated reference frame. The WEP is then verified to hold for NCQM at least as long as we consider that the noncommutative parameters are isotropic. Hence, bounds on the WEP turn out to be limits on the isotropy of the NC parameters.

Finally, the phase difference between the wavefunctions $\Phi$ and $\Phi'$ is given by:
\begin{equation}
\Psi=\mathrm{e}^{\mathrm{i}\left(\frac{m}{\hbar}\frac{\mathrm{d}\sigma}{\mathrm{d}t}x-\frac{\eta}{2\hbar^2}\sigma(t)y+\mu(t)\right)}\Psi'
\end{equation}
and, as it has been analysed in Ref. \cite{Bastos:2008b}, this does not give rise to any physically meaningful effect.

\subsection{Anisotropic noncommutativity}

As we have seen in the last subsection, the WEP holds in NCQM, unless NC parameters are anisotropic, i.e. $\eta_{xy}\neq\eta_{xz}$. In what follows we use the bounds on the WEP to constrain the difference between components of the $\eta$ matrix. The ensued discussion is similar to the one carried out in Ref. \cite{Bastos:2010au} in the context of the entropic gravity proposal \cite{Verlinde:2010hp}.
The noncommutative Hamiltonian for the GQW is given by Eq. (\ref{NCGQWH}).
In order to find the eigenstates for this problem we use perturbation theory up to first order in $\eta$, which is sufficient to obtain differences in the energy spectrum for different directions of the gravitational field. For this purpose we define
\begin{equation}
\hat{H}^{NC}=\hat{H}_0^{NC}+\hat{V},
\end{equation}
where we consider $\hat{V}$ a perturbation to the exactly soluble Hamiltonian $\hat{H}_0^{NC}$, defined by

\begin{subequations}
\begin{equation}
\hat{H}_0^{NC}:=\frac{\hat{p}_x}{2m}+\frac{\hat{p}_y}{2m}+mg\hat{x},
\end{equation}
\begin{equation} \label{perturbation}
\hat{V}:=\frac{\eta}{2m\hbar}\left(\hat{y}\hat{p}_x-\hat{x}\hat{p}_y\right)+\frac{\eta^2}{8m\hbar^2}\left(\hat{x}^2+\hat{y}^2\right).
\end{equation}
\end{subequations}

Since we are only interested in the corrections of order $\eta$, we can disregard the second term in $\hat{V}$. The soluble Hamiltonian is that of a free particle in the $y$ direction and that of the GQW in the $x$ direction. Solutions to these problems are well-known and are given by (e.g. Ref. \cite{Landau})

\begin{equation} \label{solution}
\Psi_{nk}(x,y)=A_nAi\left(\left(\frac{2m^2g}{\hbar^2}\right)^{1/3}\left(x-\frac{E_{n}}{mg}\right)\right)\chi(y),
\end{equation}
where $Ai(z)$ is the Airy function, $\chi(y)$ is the solution for the free particle, and $E_{n}$ and $A_n$ are the energy eigenvalues in the $x$ direction and the normalization factor for the Airy function, given, respectively, by,

\begin{equation}
E_{n}=-\left(\frac{mg^2\hbar^2}{2}\right)^{1/3}\alpha_n,
\end{equation}

\begin{equation}
A_n=\left[\left(\frac{\hbar^2}{2m^2g}\right)^{1/3}\int_{\alpha_n}^{+\infty}\mathrm{d}zAi^2(z)\right]^{-1/2},
\end{equation}
where $\alpha_n$ are the zeros of the Airy function. The energy eigenvalues in the $y$ direction are given by,
\begin{equation}
E_{y}=\frac{\hbar^2k^2}{2m},
\end{equation}
where $k$ is the momentum of the particle. The change in energy is given by the expectation value of the operator $\hat{V}$ in a general state given by Eq. (\ref{solution}) and, the leading order perturbation to the energy of the system in any state, is given by,

\begin{equation}
\Delta E_n=\bra{\Psi_{nk}}\hat{V}\ket{\Psi_{nk}}=\frac{\eta k}{2m}\left[\left(\frac{2m^2g}{\hbar^2}\right)^{-2/3}\mathrm{I}_1^{(n)}+\frac{E_n}{mg}\right].
\end{equation}
It must be noted that we computed the energy eigenvalues for the case of a two dimensional Hamiltonian in the $xy$ plane, so we can write,
\begin{equation}
E_{nk}^{xy}=-\left(\frac{mg^2\hbar^2}{2}\right)^{1/3}\alpha_n+\frac{\hbar^2k^2}{2m}+\frac{\eta_{xy} k}{2m}\left[\left(\frac{2m^2g}{\hbar^2}\right)^{-2/3}\mathrm{I}_1^{(n)}+\frac{E_n}{mg}\right].
\end{equation}
Thus an anisotropy in the momentum space breaks the equivalence principle. 

Consider now the NC GQW for a particle moving along the $y$ direction with a gravitational field in the $x$ direction and the same equation for a particle traveling along the $x$ direction with a gravitational field in the $z$ direction. Assuming that the test particles have the same momentum in the direction in which they are free, hence:

\begin{equation} \label{deltag}
mx(g_x-g_z)=\frac{k}{2m}\left[\left(\frac{2m^2g}{\hbar^2}\right)^{2/3}\mathrm{I}_1^{(n)}+\frac{E_n}{mg}\right]\left(\eta_{xy}-\eta_{yz}\right),
\end{equation}
where $x$ is the position of the test particle. Thus, using the bound on the WEP for two different directions (see e.g. Ref. \cite{PhysRevLett.100.041101}):

\begin{equation} \label{torsion balance}
\frac{\Delta a}{a}:=\frac{|a_1-a_2|}{a}\lesssim 10^{-13},
\end{equation}
plus data from Ref. \cite{Nesvizhesky} , namely that $k=1.03\times 10^8\>\>m^{-1}$ and $x=12.2\>\>\mu m$ for the eigenstate of lower energy and $g=9.80665\>\>m/s^2$, Eq. (\ref{deltag}) yields:

\begin{equation} \label{relation}
\frac{\Delta g}{g}=1.4\times 10^{60}\Delta\eta.
\end{equation}

Applying the bound from Eq. (\ref{torsion balance}) to Eq. (\ref{relation}), the bound for $\Delta\eta$ is computed to be:

\begin{equation}
\Delta\eta\lesssim 10^{-73} \>\mathrm{kg}^2\mathrm{m}^2\mathrm{s}^{-2},
\end{equation}
which bounds the noncommutative momentum anisotropy in a quite stringent way. In natural units:

\begin{equation}
\sqrt{\Delta\eta}\lesssim10^{-10}\>\>\mathrm{eV}.
\end{equation}

\section{Lorentz invariance}
Lorentz symmetry is a fundamental cornerstone of all known physical theories. Thus, it is natural to consider experimental bounds on this invariance to constrain noncommutativity which explicitly violates Lorentz symmetry . A major tool for these tests is the relativistic dispersion relation,
\begin{equation} \label{dispersion relation}
E^2=p^2c^2+m^2c^4.
\end{equation} 

This relation is tested with great accuracy at very high energies. Indeed, ultra-high energy cosmic rays allow for constraining this relationship for an extra quadratic term on the energy to the $1.7\times 10^{-25}$ level \cite{Bertolami:1999da}. This estimate is confirmed through direct measurements by the Auger collaboration \cite{Auger}. 

Thus, assuming a correction of the dispersion relation proportional to $E^2$ at the $1.7\times 10^{-25}$ level \cite{Bertolami:1999da}, then it is possible to constrain the $\eta$ parameter, that is:  
\begin{equation}
\eta \leqslant (1.7\times 10^{-25}) E^2,
\end{equation}
hence for ultra-high energy cosmic rays, with $E\sim 10^{20} \> \mathrm{eV}$, we can establish that $\sqrt{\eta}\leqslant 4.1\times 10^{7}\>\mathrm{eV}$, which is not at all a very stringent upper bound. A much more constraining bound can be set through low-energy tests of Lorentz symmetry. Indeed, assuming limits arising from the nuclear Zeeman levels, one can establish that $\eta \leqslant  10^{-22}E^2$, which for $E \sim \mathrm{MeV}$ \cite{PhysRevLett.57.3125}, implies that $\sqrt{\eta}\leqslant 10^{-11}\>\mathrm{MeV}\simeq 10^{-5}\>\>\mathrm{eV}$. This result is competitive with the most stringent bound on $\eta$, namely $\sqrt{\eta}\leqslant 2 \times 10^{-5}\>\>\mathrm{eV}$ \cite{Bertolami:2011rv}, obtained from the hydrogen hyperfine transition, the most accurate experimental result ever obtained.

\section{Discussion and Conclusions}

In this work we have addressed several issues on NCQM. Gauge invariance of the electromagnetic field is verified to hold only if the parameter $\theta$ vanishes, which is consistent with previous work for fermionic fields \cite{Bertolami:2011rv}. This result implies that, for abelian gauge theories, spatial directions do commute and noncommutative effects are expected only for the momenta.

Also, we have compared the GQW Hamiltonian in the context of NCQM with the Hamiltonian for the same problem in QM. Using the Fock space formalism with creation and annihilation operators, we found no evidence for a connection between this two problems at first order in the parameter $\eta$. This shows that NCQM poses a different problem from QM at least in the context of GQW.
Following this result, we studied the WEP in the noncommutative scenario. It is concluded that this principle holds for NCQM in the sense that an accelerated frame of reference is locally equivalent to a gravitational field, as long as noncommutativity is isotropic. If an anisotropy is introduced in the noncommutative parameters, using data from Refs. \cite{PhysRevLett.100.041101,Nesvizhesky}, we set a bound on the anisotropy of the $\eta$ parameter, $\sqrt{\Delta\eta}\lesssim10^{-10}\>\>\mathrm{eV}$. It is then clear that the anisotropy of the noncommutative momentum parameter is many orders of magnitude smaller than the NC parameter itself. This result also states that the existence of a preferential observer to whom the spatial $x$,$y$ and $z$ directions are well defined is limited to the same degree as the anisotropy factor.

Additionally, the breaking of Lorentz symmetry is examined in the context of NCQM. Assuming a violation of the relativistic dispersion relation proportional to $E^2$, bounds from ultra-high energy cosmic rays (see Refs. \cite{Bertolami:1999da,Auger}) imply that $\sqrt{\eta}\leq 4.1\times 10^7\>\>\mathrm{eV}$. Considering instead bounds arising from nuclear Zeeman levels, one can obtain that $\sqrt{\eta}\leq 10^{-5}\>\>\mathrm{eV}$, which is competitive with bounds arising from the hydrogen hyperfine transition $\sqrt{\eta}\leq 2\times 10^{-6}\>\>\mathrm{eV}$ \cite{Bertolami:2011rv}, the most stringent bound ever obtained.  

\vspace{0.5cm} 

\noindent
{ \bf Acknowledgements}

\noindent
The authors would like to thank Catarina Bastos for relevant discussions on the matter of this work.


\begin{thebibliography}{99}

\bibitem{Nair:2001}
V. P. Nair and A. P. Polychronakos, \textit{Phys. Lett.} B \textbf{505}, 267 (2001).

\bibitem{Duval:2001}
C. Duval and P. A. Horvathy, \textit{Journ. Phys.} A \textbf{34}, 10097 (2001).

\bibitem{Gamboa:2001}
J. Gamboa, M. Loewe and J. C. Rojas, \textit{Phys. Rev.} D \textbf{64}, 067901 (2001).

\bibitem{Pei-Ming:2002}
Pei-Ming Ho and Hsien-Chung Kao, \textit{Phys. Rev. Lett.} \textbf{88}, 151602 (2002).

\bibitem{Horvathy:2002}
P. A. Horvathy, \textit{Ann. Phys.} (N. Y.) \textbf{299}, 128 (2002).

\bibitem{Zhang:2004}
Jian-zu Zhang, \textit{Phys. Rev. Lett.} \textbf{93} 043002 (2004); \textit{Phys. Lett.} B \textbf{584}, 204 (2004).

\bibitem{Acatrinei:2005}
C. Acatrinei, \textit{Mod. Phys. Lett.} A \textbf{20}, 1437 (2005).

\bibitem{Bertolami:2005jw}
O. Bertolami, J.G. Rosa, C.M.L. de Arag\~{a}o, P. Castorina, and D. Zappal\a. \textit{Phys. Rev.} D \textbf{72}, 025010 (2005); \textit{Mod. Phys. Lett.} A \textbf{21}, 795-802 (2006).

\bibitem{Bastos:2006ps}
C. Bastos, O. Bertolami, N. C. Dias, and J. N. Prata. \textit{J. Math. Phys.} \textbf{49}, 072101 (2008).

\bibitem{2010CMaPh.299..709B}
C. Bastos, N. C. Dias, and J. N. Prata. \textit{Comm. Math. Phys.} \textbf{299}, 709-740 (2010).


\bibitem{Bernardini}
A. E. Bernardini and O. Bertolami, \textit{Phys.Rev.} A \textbf{88}, 012101 (2013).

\bibitem{SW}
N. Seiberg and E. Witten, \textit{JHEP} \textbf{9909}, 032 (1999).

\bibitem{Connes}
A. Connes, M. R. Douglas and A. Schwarz, \textit{JHEP} \textbf{02}, 003 (1998).

\bibitem{BastosQCBH}
C. Bastos, O. Bertolami, N. C. Dias and J. N. Prata, \textit{Phys. Rev.} \textbf{D78}, 023516 (2008); \textbf{D80}, 124038 (2009); \textbf{D82}, 041502 (2010); D \textbf{84}, 024005 (2011).

\bibitem{Bertolami:2011rv}
O. Bertolami and R. Queiroz, \textit{Phys. Lett.} A \textbf{375}, 4116-4119 (2011).

\bibitem{Landau}
L.D. Landau and E.M. Lifshitz, \textit{Quantum Mechanics} (Pergamon Press, 1965).

\bibitem{LF}
V.I. Luschikov and A.I. Frank, \textit{JEPT} \textbf{28}, 559 (1978).

\bibitem{Nesvizhesky}
V. V. Nesvizhesky et al., \textit{Nature} \textbf{415}, 297 (2002); \textit{Phys. Rev.} D \textbf{67} 102002 (2003).

\bibitem{Bastos:2010au}
C. Bastos, O. Bertolami, N. C. Dias, and J. N. Prata, \textit{Class. Quant. Grav.} \textbf{28}, 125007 (2011).

\bibitem{Bertolami:2003}
O. Bertolami and F. M. Nunes, \textit{Class. Quant. Grav.} \textbf{20}, L61 (2003).

\bibitem{Bertolami:2012}
O. Bertolami and J. P\'{a}ramos, in \textit{Springer Handbook of Spacetime}, Eds. Abhay Ashtekar and Vesselin Petkov, (Springer Verlag 2014).

\bibitem{Bastos:2008b}
C. Bastos and O. Bertolami, \textit{Phys. Lett.} A \textbf{372}, 5556-5559 (2008).

\bibitem{Verlinde:2010hp}
E. P. Verlinde. \textit{JHEP} \textbf{1104}, 029 (2011).

\bibitem{PhysRevLett.100.041101}
S. Schlamminger, K.-Y. Choi, T. A. Wagner, J. H. Gundlach, and E. G. Adelberger, \textit{Phys. Rev. Lett.} \textbf{100}, 041101, (2008).

\bibitem{Bertolami:1999da}
O. Bertolami and C.S. Carvalho, \textit{Phys. Rev.} D \textbf{61}, 103002 (2000).

\bibitem{Auger} F.W. Stecker and S.T. Scully, {\it New J. Phys.} {\bf 11}, 085003 (2009).

\bibitem{PhysRevLett.57.3125}
S. K. Lamoreaux, J. P. Jacobs, B. R. Heckel, F. J. Raab, and E. N. Fortson, \textit{Phys. Rev. Lett.} \textbf{57}, 3125-3128 (1986).



















\end{thebibliography}

\end{document}